\documentclass[useAMS,usenatbib]{mn2e}
\usepackage{psfig, epsf, epsfig}

\title[A ULX associated with a cloud collision in M\,99]
{A ULX associated with a cloud collision in M\,99}
\author[R. Soria \& D. S. Wong]
{Roberto Soria$^{1,2}$\thanks{E-mail:
rsoria@cfa.harvard.edu} and Diane Sonya Wong$^{3}$\thanks{E-mail:
dianew@astron.berkeley.edu}\\
$^{1}$\/Harvard-Smithsonian Center for Astrophysics, 
	60 Garden st, Cambridge, MA 02138, USA\\
$^{2}$\/Mullard Space Science Laboratory (UCL), Holmbury St Mary, 
	Dorking, Surrey, RH5 6NT, UK\\
$^{3}$\/Astronomy Department, 601 Campbell Hall, University 
	of California at Berkeley, CA 94720-3411, USA}

\begin{document}

\date{Received 11 August 2006; accepted 24 August 2006}

\pagerange{\pageref{firstpage}--\pageref{lastpage}} \pubyear{2006}

\maketitle

\label{firstpage}

\begin{abstract}
The Sc galaxy M\,99 in the Virgo cluster has been strongly 
affected by tidal interactions and recent close encounters, 
responsible for an asymmetric spiral pattern and a high star 
formation rate. Our {\it XMM-Newton} study shows that the inner disk is 
dominated by hot plasma at $kT \approx 0.30$ keV, with a total 
X-ray luminosity $\approx 10^{41}$ erg s$^{-1}$ in the 
$0.3$--$12$ keV band. At the outskirts of the galaxy, away from 
the main star-forming regions, 
there is an ultraluminous X-ray source (ULX) with an X-ray luminosity 
$\approx 2 \times 10^{40}$ erg s$^{-1}$ and a hard spectrum 
well fitted by a power law of photon index 
$\Gamma \approx 1.7$. This source is close to the location 
where a massive H\,{\footnotesize I} cloud appears to be 
falling onto the M\,99 disk at a relative speed $> 100$ km s$^{-1}$. 
We suggest that there may be a direct 
physical link between fast cloud collisions 
and the formation of bright ULXs, which may be powered 
by accreting black holes with masses $\sim 100 M_{\odot}$. 
External collisions may trigger large-scale dynamical collapses 
of protoclusters, leading to the formation of very massive 
($\ga 200 M_{\odot}$) stellar progenitors; we argue that 
such stars may later collapse into massive black holes 
if their metal abundance is sufficiently low. 
\end{abstract}


\section{Introduction}

The Sc galaxy M\,99 (NGC\,4254), 
located a distance of about $17$ Mpc (Tully 1988), 
is the brightest spiral in the Virgo Cluster ($M_B = -20.8$), 
with a gas mass $\approx 5 \times 10^{9} M_{\odot}$ 
and a kinematic mass $\approx 10^{11} M_{\odot}$ 
(Vollmer, Huchtmeier \& van Driel 2005; 
Phookun, Vogel \& Mundi 1993). 
%m_FIR = -22
It shows a peculiar spiral structure\footnote{As 
a matter of historical curiosity, it was the second galaxy 
ever in which a spiral pattern was identified (Rosse 1850), 
with the ``Leviathan of Parsonstown'', a few 
months after a similar discovery for M\,51.}, with one arm 
less tightly wound and much brighter than the other two.
%Rosse W. P., 1850, Phil. Trans. Royal Soc. London, 140, 499
Another unusual feature of M\,99 is the 
presence of a string of H\,{\footnotesize I} ``blobs'' 
to the south and west of the stellar disk, 
and a low-surface-density H\,{\footnotesize I} gas tail 
to the north-west; the total gas mass in those extra-disc 
structures is  $\approx 2 \times 10^{8} M_{\odot}$ (Phookun et al.~1993).
It is likely that there is a physical connection between 
the unusual spiral structure and the disturbed gas distribution.
In one scenario (Vollmer et al.~2005), 
the spiral structure was affected   
by a close encounter with another Virgo Cluster galaxy 
$\approx 280$ Myr ago; the surrounding H\,{\footnotesize I} clouds
are due to ongoing, face-on ram-pressure stripping, 
and are mostly moving away from the galaxy.
An alternative scenario (Phookun et al.~1993) suggests 
that the H\,{\footnotesize I} clouds and tail are tidal debris 
of an entity that was at least partly disrupted
in a close encounter with M\,99; that debris is now 
infalling and merging with the galactic disk. Simulations show 
(Bekki, Koribalski \& Kilborn 2005) that tidal debris
could also have been stripped from the outer disk of M\,99 
itself, by the Virgo Cluster potential 
as the galaxy crossed the central region of the cluster;
part of the stripped gas would later fall back onto the same galaxy. 
In this scenario, heavy gas infall onto the M\,99 disk would 
be responsible for the lopsided spiral structure 
(Phookun et al.~1993; Bournaud et al.~2005).

Intriguingly, a large H\,{\footnotesize I} cloud (VIRGOHI\,21), 
with a gas mass $\approx 2 \times 10^8 M_{\odot}$ but without 
any associated stars, located $\approx 120$ kpc to the north-west 
of M\,99, has recently been discovered (Davies et al.~2004). 
One interpretation (Minchin et al.~2005, 2006) is that it is 
an old, bound ``dark galaxy'' that has never formed stars 
because of its low hydrogen surface density. In this scenario, 
VIRGOHI\,21 would be dark-matter dominated, 
with a kinematic mass $\approx 10^{11} M_{\odot}$;
if so, it would likely be the same galaxy responsible for 
the suspected close encounter with M\,99 and 
possible tidal gas stripping.
Alternatively, VIRGOHI\,21 could itself be simply another, 
larger piece of tidal debris (without a dark matter halo), 
stripped from the outer H\,{\footnotesize I} disk of M\,99 
by the Virgo Cluster potential or during 
a close encounter with another galaxy (Bekki et al.~2005)

Gas-rich galaxies with recent tidal interactions 
generally show particularly active star formation, 
and M\,99 is no exception, with a rate 
$\approx 10 M_{\odot}$ yr$^{-1}$ inferred from 
its H$\alpha$ luminosity (Kennicutt et al.~2003), 
or $\approx 10$--$20 M_{\odot}$ yr$^{-1}$ 
from an optical/near-IR photometric study 
(Gonzales \& Graham 1996).
%Gonzalez, Rosa A.; Graham, James R. ApJ 460, p.651
As a comparison, this is twice as high as in M\,82 
(Kennicutt et al.~2003), and 
three times the current total rate 
in our Local Group (Hopkins, Irwin \& Connolly 2001).
Another phenomenon often associated with tidal interactions 
and high star formation is the presence of ultraluminous 
X-ray sources (ULXs) with apparent isotropic X-ray 
luminosities $\sim 10^{39}$--$10^{40}$ erg s$^{-1}$ 
(for recent reviews, see King 2006;
Fabbiano \& White 2006; Colbert \& Miller 2004). 
They are interpreted as accreting binary systems, 
an order of magnitude more luminous than 
the Eddington limit for typical stellar-mass 
black holes (BHs) in the Local Group.
Two ULXs were detected by {\it ROSAT}/HRI within the $D_{25}$ region 
of M\,99, on 1997 June 30 (Colbert \& Ptak 2002; Liu \& Bregman 2005).
In this paper, we use {\it XMM-Newton} to have a better 
understanding of the X-ray properties of M\,99, 
and in particular of its brightest ULX. We compare 
radio and X-ray information to investigate 
the possible connection between ULX formation, 
star formation and collisional events in this galaxy.
Finally, we outline a possible general scenario 
for the nature of ULXs, to be tested by future studies.

\section{{\it XMM-Newton} study}

\subsection{Data analysis}

M\,99 was observed with all instruments on-board {\it XMM-Newton} 
on 2003 June 26 (Revolution 651; observation ID 0147610101); 
the thin filter, full-frame mode 
was used for the European Photon Imaging Camera (EPIC) detectors. 
We processed the Observation Data
Files with standard tasks in the Science Analysis System 
({\small SAS}), version 6.5.0. Unfortunately, most of the X-ray observation 
was disrupted by strong solar flares: after inspecting 
the background lightcurves, we retained a good live-time interval 
of 17.0 ks for the pn and 20.5 ks for the MOSs, 
out of the total 51-ks exposure. In addition, 
Optical Monitor (OM) images were taken in the $UVW1$, 
$UVM2$ and $UVW2$ filters, with exposure times 
of $3.0$, $6.0$ and $17.8$ ks, respectively\footnote{See 
http://www.xmm.ac.uk/onlines/uhb/XMM\_UHB/node66.html 
for a plot of the OM filter throughput curves.}.
We filtered the EPIC event files, selecting only the best-calibrated 
events (pattern $\leq$ 12 for the MOSs, pattern $\leq$ 4 for pn), 
and rejecting flagged events. 

We used the Chandra Interactive Analysis 
of Observations ({\small CIAO}) task {\tt wavdetect} to identify 
discrete sources in the combined EPIC image, 
in different energy bands\footnote{A comparison 
between the source-finding algorithms 
in {\small CIAO} and {\small SAS} can be found in 
Valtchanov, Pierre \& Gastaud (2001).}. We selected 
a source extraction region of radius $25\arcsec$ around 
the brightest point-like source (the only one of the two 
{\it ROSAT} ULXs that is seen again by {\it XMM-Newton}, 
see Section 2.2). We built response functions with the {\small SAS} 
task {\tt rmfgen}, and auxiliary response functions with {\tt
arfgen}, for pn and the two MOSs. We then coadded 
the three spectra, with the method described in Page, Davis \& Salvi (2003), 
in order to increase the signal-to-noise ratio. 
Similarly, we obtained combined EPIC spectra 
of the X-ray emission from the whole galaxy (excluding the ULX) 
and from the star-forming region within $1\arcmin$ from 
the galactic centre. 
Finally, we fitted the background-subtracted spectra 
with standard models in {\small XSPEC} version 11.3.1 
(Arnaud 1996).

\subsection{Discrete X-ray sources and unresolved emission}

The X-ray emission from M\,99 is dominated 
by hot gas, as expected from its high star formation rate (SFR). 
The modest spatial resolution 
of {\it XMM-Newton} and the strength of the diffuse component
make it difficult to resolve individual point-like 
sources (essentially, high-mass X-ray binaries 
and young supernova remnants), especially in the inner 
arcmin ($4.8$ kpc at the assumed distance), 
where most of the X-ray emission is concentrated 
(Figure 1). A few point-like sources are instead 
detected (Table 1) from the combined EPIC image in the 
$1.5$--$12$ keV band, where 
the diffuse emission is negligible\footnote{The same five 
sources listed in Table 1 are also found when we use 
the SAS source-finding routine {\tt emldetect}.}. 
The location of the detected sources is shown 
in Figure 2, overplotted on a true-colour OM image.
Of the two ULXs found by {\it ROSAT} inside the $D_{25}$ 
ellipse in 1997, one (X-1 in the {\it ROSAT} catalogue, 
Liu \& Bregman 2005)
is no longer detected: hence, we estimate that it must 
now be fainter than $\approx 5 \times 10^{38}$ erg s$^{-1}$ 
in the $0.3$--$12$ keV band; its luminosity 
was $\approx 6 \times 10^{39}$ erg s$^{-1}$ in 1997,
extrapolated to the same band.
The other ULX (IXO\,46 in Colbert \& Ptak 2002, 
or X-2 in Liu \& Bregman 2005) is instead 
recovered, with a $0.3$--$12$ keV luminosity 
$\approx 2 \times 10^{40}$ erg s$^{-1}$ 
(detailed analysis of this source in Section 2.4).

\begin{table*}
         \begin{tabular}{lccccc}
            \hline
            \noalign{\smallskip}
No.
%& {\mathrm{CXOU~Name}} 
& R.A.(2000)     
& Dec.(2000)
                 & $f_{1.5-12}$
                 & $f_{0.3-12}$
                 & $L_{0.3-12}$\\
%                 & Notes \\
&&&($10^{-14}$ erg cm$^{-2}$ s$^{-1}$)& ($10^{-14}$ erg cm$^{-2}$ s$^{-1}$) 
& ($10^{39}$ erg s$^{-1}$)  \\
            \noalign{\smallskip}
            \hline
            \noalign{\smallskip}
01 & $12~18~45.58$ & $14~26~43.1$ 
   & $1.71\pm0.35$ & $2.46\pm0.50$ & $0.85\pm 0.20$ \\[3pt]
02 & $12~18~51.13$ & $14~26~30.4$ 
   & $2.09\pm0.38$ & $3.01\pm0.55$ & $1.0\pm 0.2$ \\[3pt]
03 & $12~18~51.36$ & $14~24~24.5$ 
   & $3.02\pm0.51$ & $4.35\pm0.73$ & $1.5\pm 0.3$ \\[3pt]
04 & $12~18~52.61$ & $14~25~47.8$ 
   & $2.44\pm0.38$ & $3.52\pm0.55$ & $1.2\pm 0.2$ \\[3pt]
05 & $12~18~56.15$ & $14~24~18.0$
   & $39.5^{+2.8}_{-1.2}$ & $56.1^{+3.4}_{-1.7}$ & $19.4^{+1.2}_{-0.6}$\\
   \noalign{\smallskip}
            \hline
         \end{tabular}
      \caption{Point-like X-ray sources  
detected at $> 3$-$\sigma$ significance in the $0.3$--$12$ keV band 
inside the $D_{25}$ ellipse of M\,99. 
Fluxes and luminosities are isotropic, emitted values, 
assuming a power-law spectrum with $\Gamma = 1.7$ 
and a line-of-sight Galactic column density 
$N_{\rm H,Gal} = 2.7 \times 10^{20}$ cm$^{-2}$.}
         \label{t:allsources}
\end{table*}

% 12 18 52.13	14 24 52.3 NOT FOUND
% < 15 counts in hard band 
% (was 4.7E39 on MJD50629, 20.8 ks)

% nucleus
% 12:18:49.774,+14:24:57.16
% 70 +/- 15
% 689 +/- 45
% ( 1.500- 12.000keV) of 2.222E-14
%  ( 0.300- 12.000keV) of 3.204E-14

We estimate that the total unabsorbed X-ray luminosity 
(not including the bright ULX) from inside the $D_{25}$ ellipse  
is $\approx (1.2 \pm 0.2) \times 10^{41}$ erg s$^{-1}$ 
in the $0.3$--$12$ keV band, of which only 
$\approx  10^{40}$ erg s$^{-1}$ in the $2$--$12$ 
keV band. About $70\%$ of the X-ray emission comes from 
within a projected radius of $1\arcmin \approx 4.8$ kpc.
The unresolved emission is dominated by thermal plasma 
at a temperature $\la 0.30$ keV; a two-temperature 
thermal-plasma fit ({\tt vapec} in {\small XSPEC}) 
gives temperatures $kT_1 = (50 \pm 5)$ eV 
and $kT_2 = (0.30 \pm 0.01)$ keV (Figure 3). 
Gas temperatures $\la 0.3$ keV are typical of disk emission 
in normal spirals, and are also one of the plasma components 
in starburst galaxies. We do not detect any additional 
thermal-plasma component at $\approx 0.7$--$0.8$ keV, which 
is instead characteristic of starbursts (e.g., Ott, Walter \& Brinks 2005; 
Jenkins et al.~2005; Summers et al.~2003, 2004).
The diffuse, soft X-ray emission, which traces recent 
star formation, appears to be uniformly distributed 
over the inner disk, and only weakly peaked in the nuclear region.

An additional, harder component of the unresolved emission 
is well fitted by a power-law of photon index 
$\Gamma = 1.70 \pm 0.25$. 
As is generally the case for star-forming galaxies, 
the power-law component can be interpreted as emission 
from unresolved X-ray binaries. The contribution of 
the discrete sources (both resolved and unresolved, 
but not including the ULX) 
to the total emitted luminosity in the $0.3$--$12$ keV 
band is only $\approx 15\%$ of the total luminosity 
inside the $D_{25}$ ellipse. 
The relative importance of the thermal-plasma emission 
over the point-source contribution 
is even higher in the inner-disk region (Table 2), where 
the power-law component is only $\approx 10\%$ of the total 
luminosity.

% MHI = 2.36 x 10^5 D^2 FHI
%Rohlfs, Wilson  2000

\begin{center}
   \begin{table}
      \caption{Best-fitting parameters 
to the combined {\it XMM-Newton}/EPIC spectrum of 
the inner-disk emission; the source region 
is a circle of radius $1\arcmin$. 
The {\small XSPEC} model is
{\tt tbabs}$_{\rm Gal}~\times$ {\tt tbabs} $\times$ ({\tt powerlaw} 
$+$ {\tt vapec}$_1$ $+$ {\tt vapec}$_2$). 
The quoted errors are the 90\% confidence limit and
$N_{\rm H,Gal} = 2.7 \times 10^{20}$ cm$^{-2}$ (Dickey \& Lockman 1990). 
$K_{\rm 1}$ and $K_{\rm 2}$ are the normalisation constants 
of the two {\tt vapec} components;
$f_{0.3{\rm -}12}^{\rm obs}$ is the observed flux; 
$f_{0.3{\rm -}12}^{\rm em}$ the unabsorbed flux; 
$(L_{\rm po}/L)_{0.3{\rm -}12}$ the relative contribution 
of the point-like accreting sources (resolved and unresolved) 
to the total X-ray emission, clearly dominated here by hot thermal plasma.} 
%We assumed a solar abundancefor the intrinsic absorber ($Z = Z_{\odot}$).}
         \label{table1c}
\begin{center}
         \begin{tabular}{lr}
%            \hline
            \hline
            \noalign{\smallskip}
            Parameter    & Value \\[2pt]
            \noalign{\smallskip}
%            \hline
            \hline
            \noalign{\smallskip}
            \noalign{\smallskip}
%        \multicolumn{3}{c}{model: wabs$_{\rm Gal}~\times$
%                wabs $\times$ po}\\
%            \noalign{\smallskip}
%            \hline
%            \noalign{\smallskip}
%            \noalign{\smallskip}
                $N_{\rm H}~(10^{20}~{\rm cm}^{-2})$ 
	                & $18.1^{+6.4}_{-1.3}$ \\[2pt]
                $\Gamma$  & $1.70^{+0.24}_{-0.23}$\\[2pt]
                $K_{\rm po}~(10^{-5})$ 
                        & $3.3^{+1.0}_{-1.2}$\\[2pt]
	        $kT_{\rm 1}~({\rm keV})$ & $0.050^{+0.05}_{-0.05}$\\[2pt]
	        $K_{\rm 1}~(10^{-3})$ & $350^{+500}_{-110}$\\[2pt]
	        $kT_{\rm 2}~({\rm keV})$ & $0.30^{+0.01}_{-0.02}$\\[2pt]
	        $K_{\rm 2}~(10^{-3})$ & $1.5^{+1.1}_{-0.2}$\\[2pt]
	        $(Z/Z_{\odot})$ & $0.13^{+0.36}_{-0.06}$\\
            \noalign{\smallskip}
            \hline
            \noalign{\smallskip}
                $\chi^2_\nu$ & $0.96 (81.0/84)$ \\[2pt] 
                $f_{0.3{\rm -}12}^{\rm obs}$
                         $~(10^{-12}~{\rm CGS})$
	                 & $0.39^{+0.02}_{-0.06}$\\[2pt]
                $f_{0.3{\rm -}12}^{\rm em}$
                        $ ~(10^{-12}~{\rm CGS})$ 
	                & $2.5^{+2.4}_{-1.5}$\\[2pt]
                $L_{0.3{\rm -}12}~(10^{40}~{\rm erg~s}^{-1})$ 
                        & $8.6^{+8.3}_{-5.2}$\\[2pt]
	        $(L_{\rm po}/L)_{0.3{\rm -}12}$ & $\approx 0.10$\\
            \noalign{\smallskip}
            \hline
         \end{tabular}
%\begin{list}{}{}
%\item[$^{\mathrm{a}}$] assumed to be equal for the two states
%\item[$^{\mathrm{b}}$] in units of $10^{-12}~{\rm erg~cm}^{-2}~{\rm s}^{-1}$
%\item[$^{\mathrm{c}}$] in units of $10^{40}~{\rm erg~s}^{-1}$
%\end{list}
\end{center}
   \end{table}
\end{center}

\subsection{Nuclear morphology: an old conundrum}

The nucleus itself is not unequivocally resolved 
as a point source. We do clearly detect X-ray emission 
in the hard band at the nuclear position (Figure 1, 
middle panel), for an estimated emitted luminosity 
$\approx 10^{39}$ erg s$^{-1}$ extrapolated 
to the $0.3$--$12$ keV band. However, because of the low number 
of counts and low spatial resolution, we cannot tell 
whether it is the true nuclear source, or instead 
a number of unresolved X-ray binaries in the region 
where star formation is more active.

On an historical note, the nature of the nuclear source 
in M\,99 was the subject of intense investigations 
by Carl Lampland at the Lowell Observatory, many decades ago 
(Lampland 1921). Lampland noted that the peak 
of the optical emission in his photographic plates 
(which have peak sensitivity at $\approx 4200$\AA, 
in the standard $B$ band) appeared sometimes 
shifted from the position corresponding (as we know today) 
to the true galactic nucleus, to another point 
located $\approx 2\farcs5$ to the southeast.
This variability was re-examined and confirmed 
by Walker (1967), who found that the off-nuclear 
source was brighter in 10 of the 30 Lowell plates 
taken between 1921 and 1948, with changes occurring 
sometimes on a timescale of a few days. Spectroscopic 
observations (Walker 1967) also suggested 
that the true optical nucleus is redder, 
corresponding to a stellar type G0--G2, 
while the off-nuclear source appears much bluer.

\begin{figure*}
\epsfig{figure=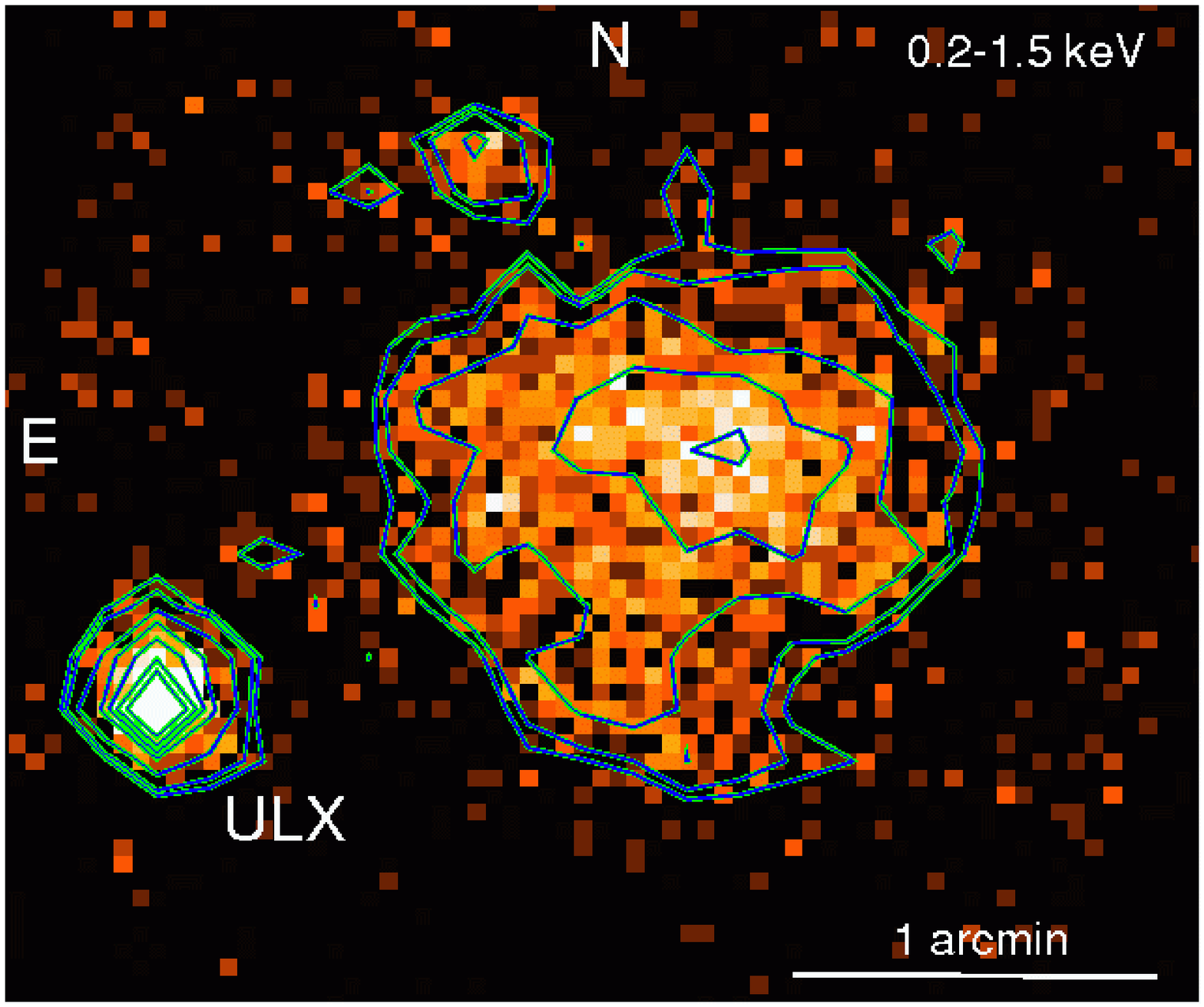, width=5.8cm, angle=0}
\epsfig{figure=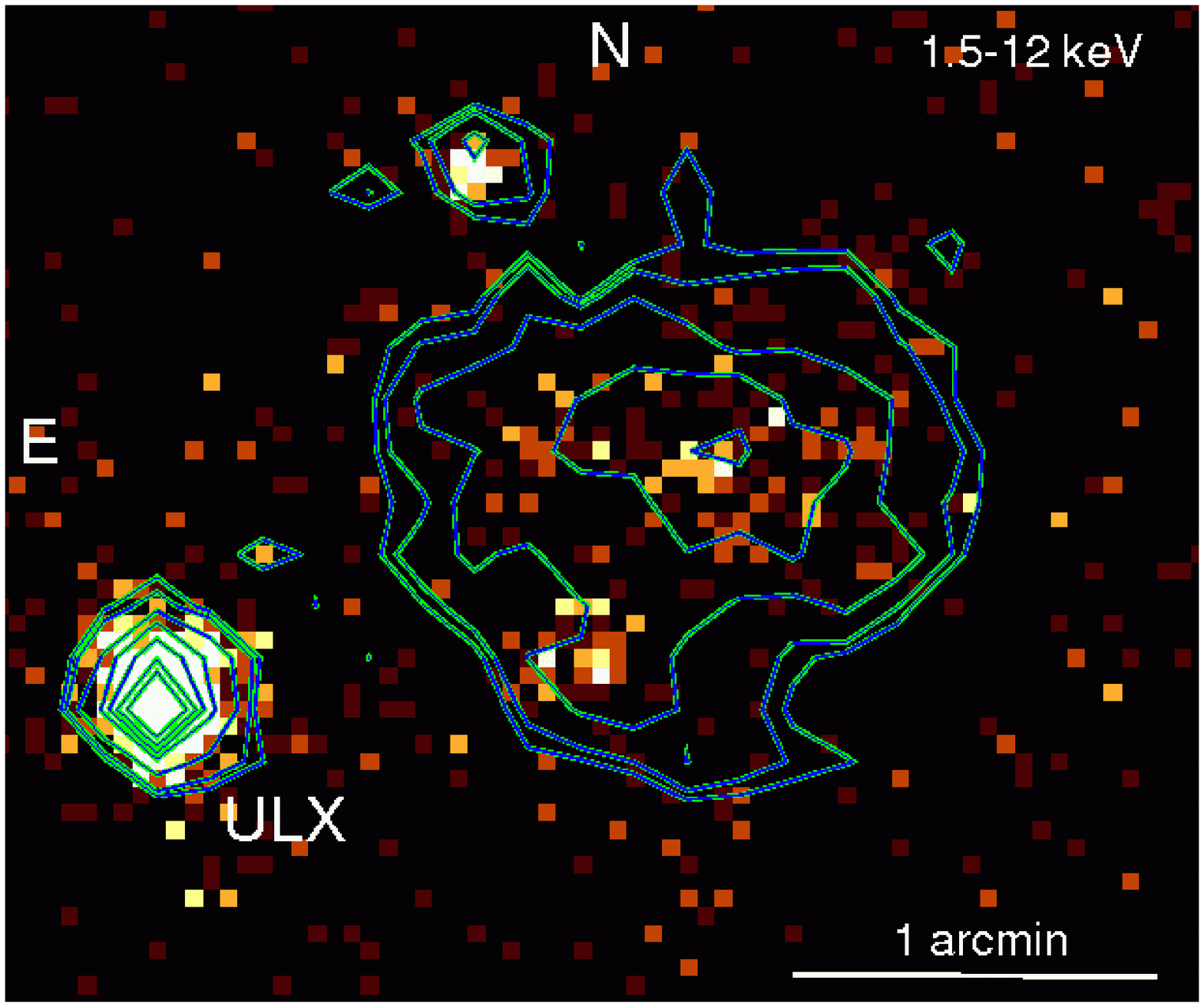, width=5.8cm, angle=0}
\epsfig{figure=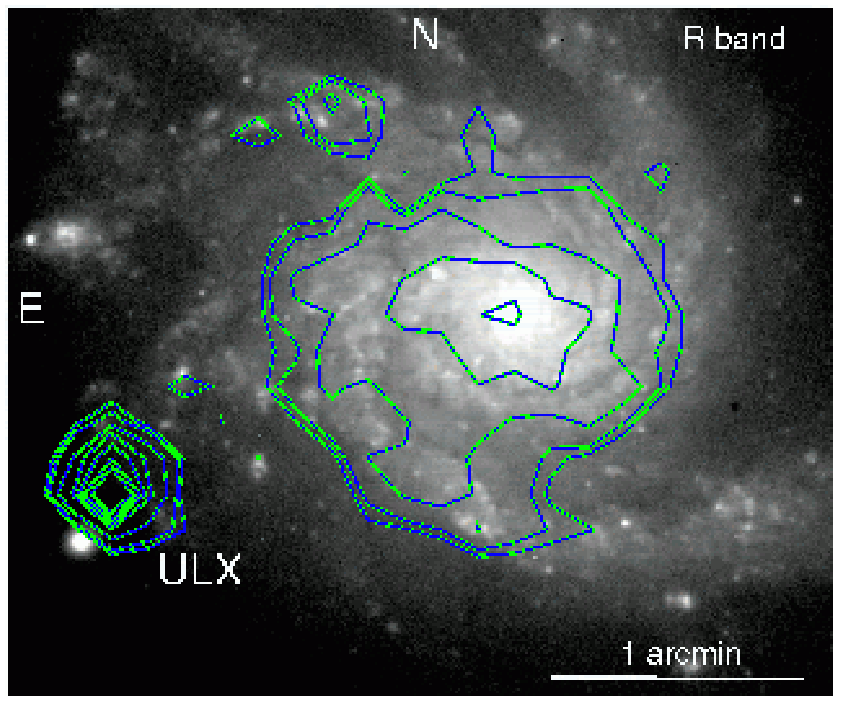, width=5.8cm, angle=0}
\caption{{\it XMM-Newton}/EPIC X-ray maps of the inner part 
of M\,99. Left panel: combined EPIC image and contours for 
the soft band ($0.2$--$1.5$ keV), dominated by 
emission from diffuse hot plasma. Middle panel: 
EPIC image in the hard band ($1.5$--$12$ keV), 
with soft-X-ray contours superimposed for comparison. 
Right panel: greyscale optical image in the Bessell-$R$ band 
(archival data from the {\it VLT}/FORS1) with soft-X-ray contours 
overplotted.}
\end{figure*}

\begin{figure}
\epsfig{figure=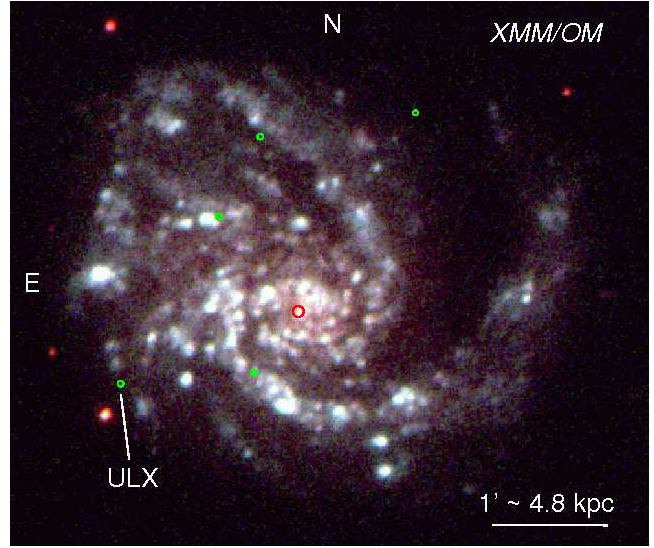, width=8.5cm, angle=0}
\caption{True-colour UV image from {\it XMM-Newton}/OM; red 
corresponds to the $UVW1$ filter, green to the $UVM2$ filter, 
and blue to the $UVW2$ filter. Point-like 
X-ray sources detected with {\it XMM-Newton}/EPIC are marked 
with green circles, of radius $1\farcs5$. 
The peak of the X-ray emission in the nuclear region 
is indicated with a red circle, of radius 3\arcsec.}
\end{figure}

We can now shed further light on the nuclear morphology.
An {\it HST}/WFPC2 image in the F606W filter 
(Figure 4, left panel) is dominated by a bright nuclear 
star cluster 
(a nuclear structure typical of late-type spiral 
galaxies), located at the dynamical centre, 
as can be inferred from the morphology of the dust 
filaments. The cluster is spatially resolved 
in the {\it HST} image, with a full width half maximum 
of $0\farcs15 \pm 0\farcs02$ corresponding to
$(12 \pm 2)$ pc; point sources in the same field 
have a width of $0\farcs10 \pm 0\farcs01$.
%$(7\pm 1)$ pc. 
An approximate conversion 
to standard colours gives an absolute magnitude 
$M_V \approx -13.5$ mag. Within $200$ pc from the nucleus, 
there are a few, fainter, unresolved stellar 
clusters with typical absolute magnitudes 
$M_V \approx -11$ mag. The secondary optical peak 
described by Lampland (1921) and Walker (1967) 
corresponds to the stellar complex labelled as A 
in Figure 4 (unresolved from ground-based telescopes), 
and sometimes (at least in plates with worse seeing) 
to the unresolved blend of A and B.
When we compare the {\it HST} image with an 
{\it XMM-Newton}/OM image of the same field, 
in the near-UV ($UVW2$ filter, with an effective 
wavelength of $2120$\AA), we note that 
the true nucleus is undetected: the brightest 
UV sources in the nuclear region are the young 
stellar complexes labelled as A, B and C 
(Figure 4, right panel). 
The optical nucleus remains undetected 
even in the $UVW1$ filter (effective 
wavelength of $2910$\AA).  
Unfortunately, the {\it XMM-Newton}/EPIC image 
does not provide enough information to determine 
whether the X-ray emission peaks at the true optical 
nucleus or at any of those three UV-bright sources nearby.
It is plausible that the true nucleus and source A 
would appear of comparable brightness 
in the standard $U$ band. The nuclear cluster 
is the brighter source in the $B$ band, as evident 
from the archival images in the NASA/IPAC Extragalactic 
Database\footnote{http://nedwww.ipac.caltech.edu/} (NED).
Of course, this does not explain how the peak 
brightness could be perceived to shift between 
the two sources several times over a few decades, in plates 
with the same density, taken a few days apart.
Repeated supernovae or variability of individual stars 
are clearly unsatisfactory explanations; 
variations in the brightness of a nuclear BH 
would also be very peculiar.
The true explanation of this effect, if real, 
remains a mystery for now.
%Further studies of this issue are beyond 
%the scope of this paper.

%Walker 1967, PASP, 79, 593. There are two ``nuclear'' 

\begin{figure}
\epsfig{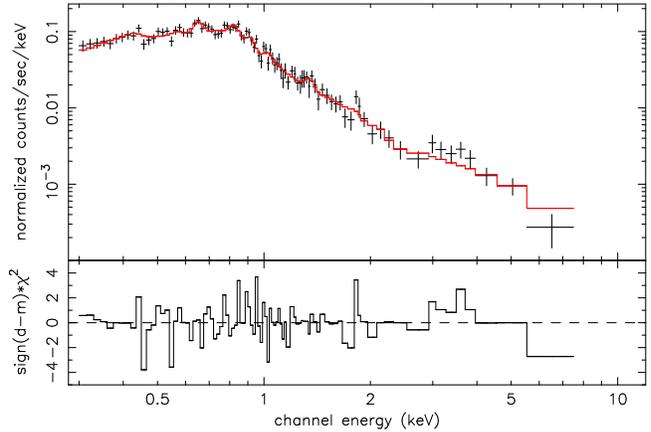}
\caption{Coadded {\it XMM-Newton}/EPIC spectrum and 
best-fit residuals of the unresolved 
emission in the inner disk (within $1\arcmin$ from the
nucleus). The spectrum has been fitted with a two-temperature 
thermal plasma component plus a power law. See Table 2 
for the best-fitting parameters.
}
\end{figure}

\begin{figure*}
\epsfig{figure=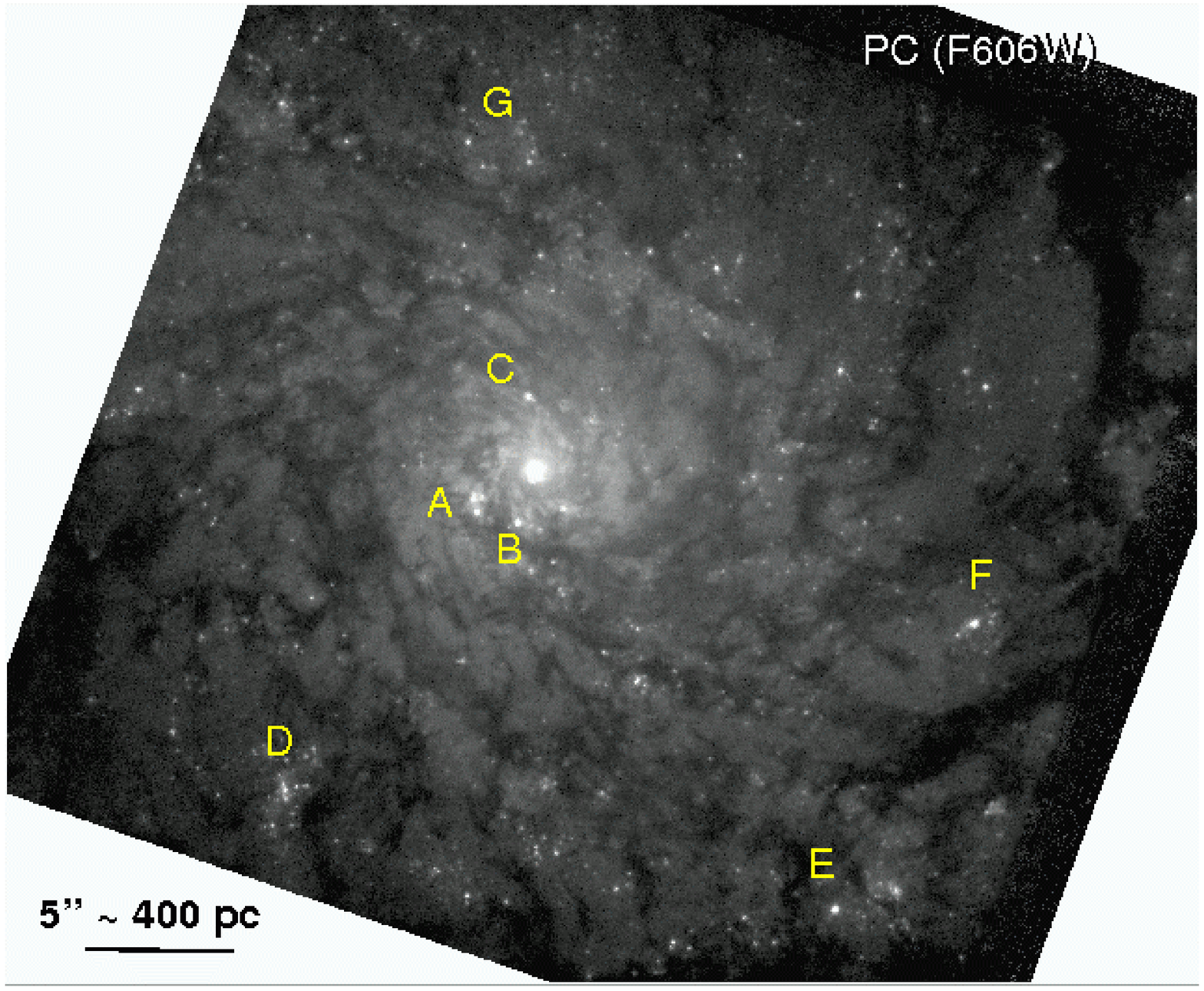, width=8.6cm}
\epsfig{figure=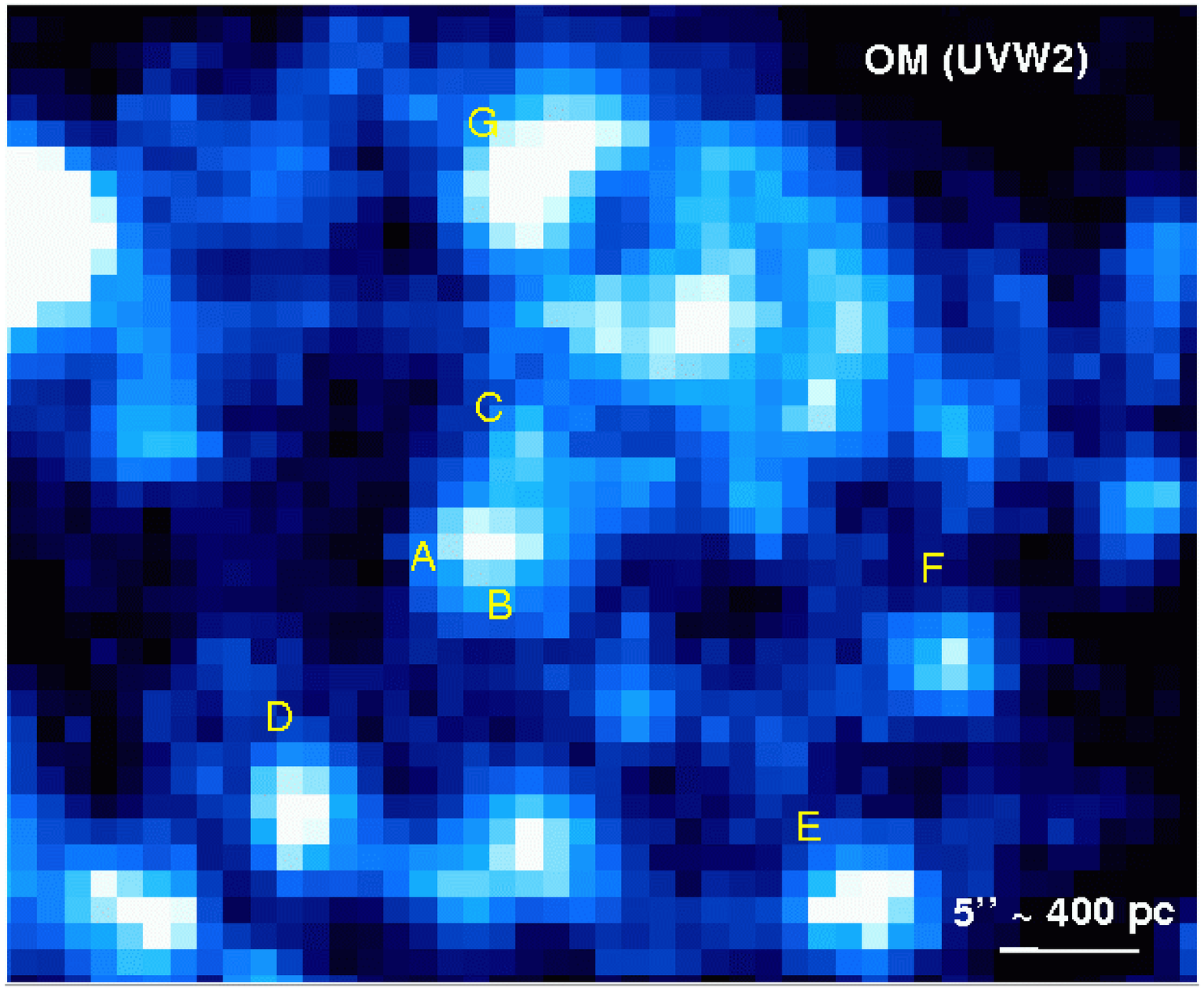, width=8.6cm}
\caption{Morphology of the nuclear region in the optical and near-UV. 
Left panel: {\it HST}/WFPC2 image in the F606W filter; right panel: 
the same region, seen by {\it XMM-Newton}/OM 
in the $UVW2$ filter. At the dynamical center, 
an old nuclear star cluster dominates the optical image 
but is undetected in the near UV. The labels identify 
smaller, younger clusters of OB stars, and are 
given simply to facilitate a comparison between 
the two images.}
\end{figure*}

\subsection{A bright, hard ULX}

The bright ULX located $\approx 8$ kpc southeast 
of the nucleus was the main objective of our X-ray study.
Above $1.5$ keV, the ULX is brighter than the rest 
of the galaxy (Figure 1). Its background-subtracted 
X-ray spectrum ($\approx 2100$ net EPIC counts) 
is well fitted by a simple absorbed 
power-law (Figure 5 and Table 3) of index $\Gamma = 1.7 \pm 0.1$. 
There is no hint of a thermal component 
(blackbody or disk-blackbody) that could be attributed 
to an accretion disk.
Its emitted luminosity in the $0.3$--$12$ keV band 
is $1.9^{+0.2}_{-0.1} \times 10^{40}$ erg s$^{-1}$, 
which is the same, within the errors, as the 
extrapolated {\it ROSAT} luminosity observed in 1997 
(Liu \& Bregman 2005).
Interestingly, the X-ray emission detected by 
{\it Einstein}/IPC (Fabbiano, Kim \& Trinchieri 1992; 
archival X-ray image available in NED) 
is centred on the galactic disc and does not 
show an enhancement at the ULX position, even allowing 
for the lower spatial resolution and low number of counts 
($140 \pm 16$). As a further comparison, 
there is another X-ray source clearly 
visible in the {\it Einstein} image, about $8\arcmin$ west 
of the galaxy (a background quasar). In the {\it XMM-Newton} 
observation, its flux in the {\it Einstein} energy band 
is less than half of the ULX flux; however, that 
background source was clearly stronger than the ULX 
in the {\it Einstein} image. We estimate that at 
the time of the {\it Einstein} 
observations (1980 June 25), the ULX must have been 
at least a factor of 5 fainter than in 1997 and 2003.
%No Einstein
%The Imaging Proportional Counter has been described by Gorenstein, et al. (TRANS. IEEE Nuc. Sci., NS-28 , 869, 1981), and by Giacconi, et al. (Ap.J. 230 , 540, 1979).

We estimate that the {\it XMM-Newton}/EPIC astrometry 
is accurate to $\approx 1\farcs5$. There are no other 
point-like X-ray sources in the field that can allow 
a better registration with optical positions. 
We searched for optical counterparts within the X-ray 
error circle in the {\it XMM-Newton}/OM images 
and in several optical images from public archives 
but found none. The bright optical source
suggested as a possible counterpart by Colbert \& Ptak (2002), 
based on the {\it ROSAT}/HRI position, 
is now clearly ruled out (Figure 1, right panel). 
Its brightness and colours are consistent with a G0 
foreground star (Kharchenko 2001). 
We took optical spectra of this 
source with the LRIS spectrograph on the Keck telescope, 
confirming the spectral identification 
and verifying that its radial velocity is consistent 
with a foreground star, ruling out the possibility 
that it belongs to M\,99. 
Using the best high-resolution optical images 
available in public archives (in particular, 
a 30-s {\it VLT}/FORS1 
observation in the Bessell-$R$ band; 
a 720-s $B$-band image from the 2.1-m telescope 
at Kitt Peak National Observatory; 
a 200-s $B$-band image from the 2.5-m 
Isaac Newton Telescope at La Palma), 
we conclude that there are no optical counterparts 
brighter than $M_B \approx -9.5$ mag 
and $M_R \approx -9$ mag. This is not 
a very stringent limit, and does not even rule out 
an old globular cluster. None the less, we can 
confidently say that the ULX 
is right at the outer edge of the stellar 
disk (the limit beyond which the gas density is 
too low to collapse and form stars spontaneously), 
and that there are no super star-clusters 
or massive OB associations at that position 
or within $\approx 800$ pc.

\begin{center}
   \begin{table}
      \caption{Best-fitting parameters 
to the combined {\it XMM-Newton}/EPIC spectrum of 
the ULX. The {\small XSPEC} model is
{\tt tbabs}$_{\rm Gal}~\times$ {\tt tbabs} $\times$ {\tt po}. 
The quoted errors are the 90\% confidence limit and
$N_{\rm H,Gal} = 2.7 \times 10^{20}$ cm$^{-2}$ (Dickey \& Lockman 1990).} 
%We assumed a solar abundancefor the intrinsic absorber ($Z = Z_{\odot}$).}
         \label{table1c}
\begin{center}
         \begin{tabular}{lr}
%            \hline
            \hline
            \noalign{\smallskip}
            Parameter    & Value \\[2pt]
            \noalign{\smallskip}
%            \hline
            \hline
            \noalign{\smallskip}
            \noalign{\smallskip}
%        \multicolumn{3}{c}{model: wabs$_{\rm Gal}~\times$
%                wabs $\times$ po}\\
%            \noalign{\smallskip}
%            \hline
%            \noalign{\smallskip}
%            \noalign{\smallskip}
                $N_{\rm H}~(10^{20}~{\rm cm}^{-2})$ 
	                & $23.3^{+3.8}_{-3.4}$ \\[2pt]
                $\Gamma$  & $1.67^{+0.10}_{-0.10}$\\[2pt]
                $K_{\rm po}~(10^{-5})$ 
                        & $7.3^{+0.9}_{-0.8}$\\
            \noalign{\smallskip}
            \hline
            \noalign{\smallskip}
                $\chi^2_\nu$ & $0.89 (86.0/97)$ \\[2pt] 
                $f_{0.3{\rm -}12}^{\rm obs}$
                         $~(10^{-13}~{\rm CGS})$
	                 & $4.4^{+0.2}_{-0.3}$\\[2pt]
                $f_{0.3{\rm -}12}^{\rm em}$
                        $ ~(10^{-13}~{\rm CGS})$ 
	                & $5.6^{+0.4}_{-0.2}$\\[2pt]
                $L_{0.3{\rm -}12}~(10^{40}~{\rm erg~s}^{-1})$ 
                        & $1.9^{+0.2}_{-0.1}$\\
            \noalign{\smallskip}
            \hline
         \end{tabular}
%\begin{list}{}{}
%\item[$^{\mathrm{a}}$] assumed to be equal for the two states
%\item[$^{\mathrm{b}}$] in units of $10^{-12}~{\rm erg~cm}^{-2}~{\rm s}^{-1}$
%\item[$^{\mathrm{c}}$] in units of $10^{40}~{\rm erg~s}^{-1}$
%\end{list}
\end{center}
   \end{table}
\end{center}

\begin{figure}
\epsfig{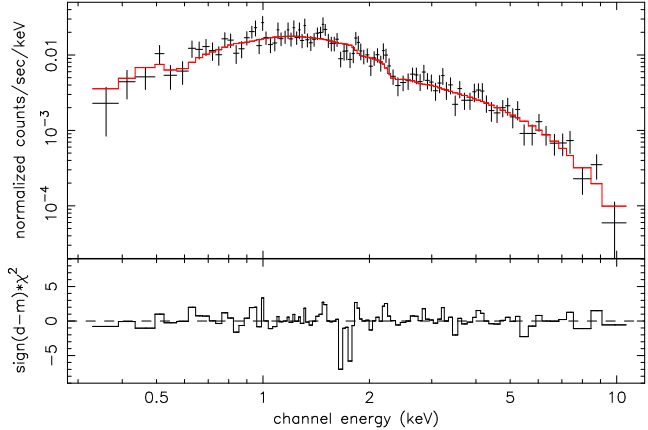}
\caption{Coadded {\it XMM-Newton}/EPIC spectrum and 
best-fit residuals of the ULX spectrum, 
fitted with a power law of index $\Gamma = 1.7\pm 0.1$.
See Table 3 for the best-fitting parameters.}
\end{figure}

\section{Radio study: a colliding cloud?}

If the optical data offer no clue on the nature of the ULX, 
the radio data may instead suggest a complex, intriguing story.
H\,{\footnotesize {I}} 21-cm line observations of M\,99 
were carried out by Phookun et al.~(1993), using 
the C and D arrays (exposure times of 8 hr and 4 hr, 
respectively) of the Very Large Array\footnote{The 
VLA is a facility of the National Radio Astronomy Observatory, 
which is operated by Associated Universities, Inc., 
under cooperative agreement with the National 
Science Foundation.} (VLA). The size of the synthesized beam 
was $25\farcs03 \times 23\farcs56$. See Phookun et al.~(1993) 
for details of the observations and data analysis.
As expected, the H\,{\footnotesize {I}} disk is $\approx 40\%$ 
larger than the stellar disk, and extends slightly 
beyond the location of the ULX (Figure 6).
The total H\,{\footnotesize {I}} flux from the galaxy 
corresponds to a gas mass $\approx 5 \times 10^9 M_{\odot}$ 
(Vollmer et al.~2005; Phookun et al.~1993), 
using a standard conversion between H\,{\footnotesize {I}} 
flux and mass (e.g., Rohlfs \& Wilson 2006).

Interestingly, the ULX is located close to where 
a large H\,{\footnotesize {I}} spur or cloud apparently joins 
the gas disk (Figure 6). The gas mass of this cloud is 
$\approx 10^7 M_{\odot}$, considering only the fraction 
that visibly ``sticks out'' from the disk. More likely, 
its mass could a factor of two higher, considering that part 
of it overlaps with the disk (see also Fig.~5 in 
Phookun et al.~1993). This cloud represents $\sim 10\%$ 
of the gas mass outside the galactic H\,{\footnotesize {I}} 
disk: either infalling onto the disk, or in the process 
of being shredded from it, according to alternative 
models (recall Section 1).  It may be possible that 
the association is simply a coincidence: there is no 
definitive way to tell from the data available. 
It is none the less remarkable that, while there are 
various other gas clouds around the galaxy 
(references in Section 1), the cloud near the ULX 
is the only one that overlaps and probably impacts 
the stellar disk; all the others are at larger 
radial distances, at the edge of the galactic 
H\,{\footnotesize {I}} disk. 
We estimate that only $\approx 3\%$ of the projected 
area of the stellar disk overlaps with this (or any other) 
external cloud.

The H\,{\footnotesize {I}} velocity map 
(Phookum et al.~1993) tells a more dramatic picture 
(Figure 7). The gas cloud is strongly blue-shifted 
with respect to the galactic disk, 
with a difference in the projected radial 
velocity $> 100$ km s$^{-1}$, and is clearly 
overlapping with the disk.
From the inferred radial velocities, the cloud could 
be either in the process of being ejected 
(ram-pressure stripped) from the disk towards us, 
or, more likely in our opinion, it is tidal debris 
infalling onto the disk from behind. In the latter case, 
we have no elements to determine whether the cloud 
is actually hitting and merging with the disk. 
But is is interesting to note that the possible 
impact point is very close to the projected ULX position.
We shall discuss in Section 4 whether there 
may be a direct physical connection between collisional 
processes and the formation of ULXs.

%{\it Summarize radio observations by Phookun (1993) 
%and show that there is an HI cloud apparently impacting
%onto the M\,99 disk at the ULX position.}

\begin{figure}
\epsfig{figure=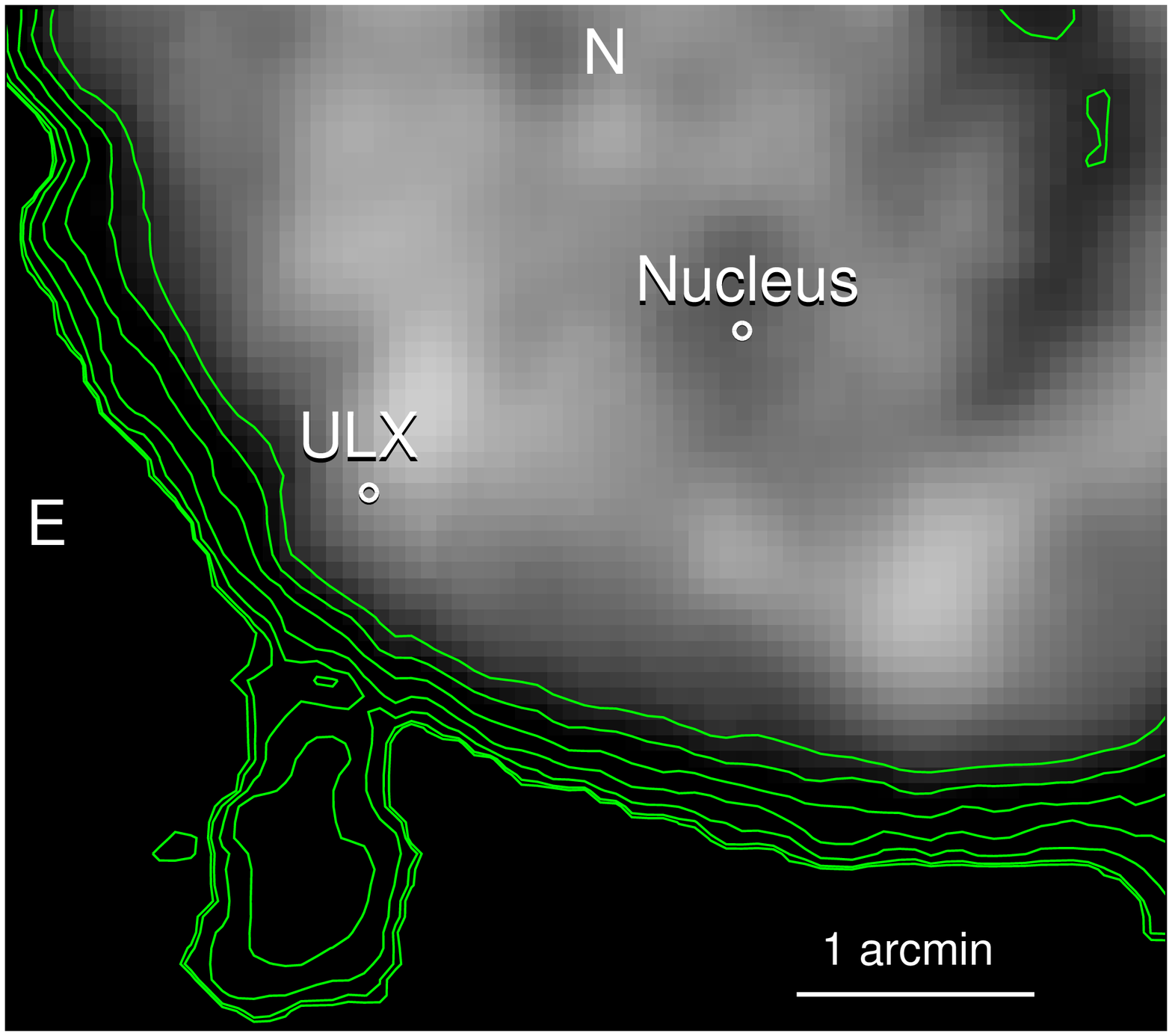, width=8.5cm, angle=0}\\
\epsfig{figure=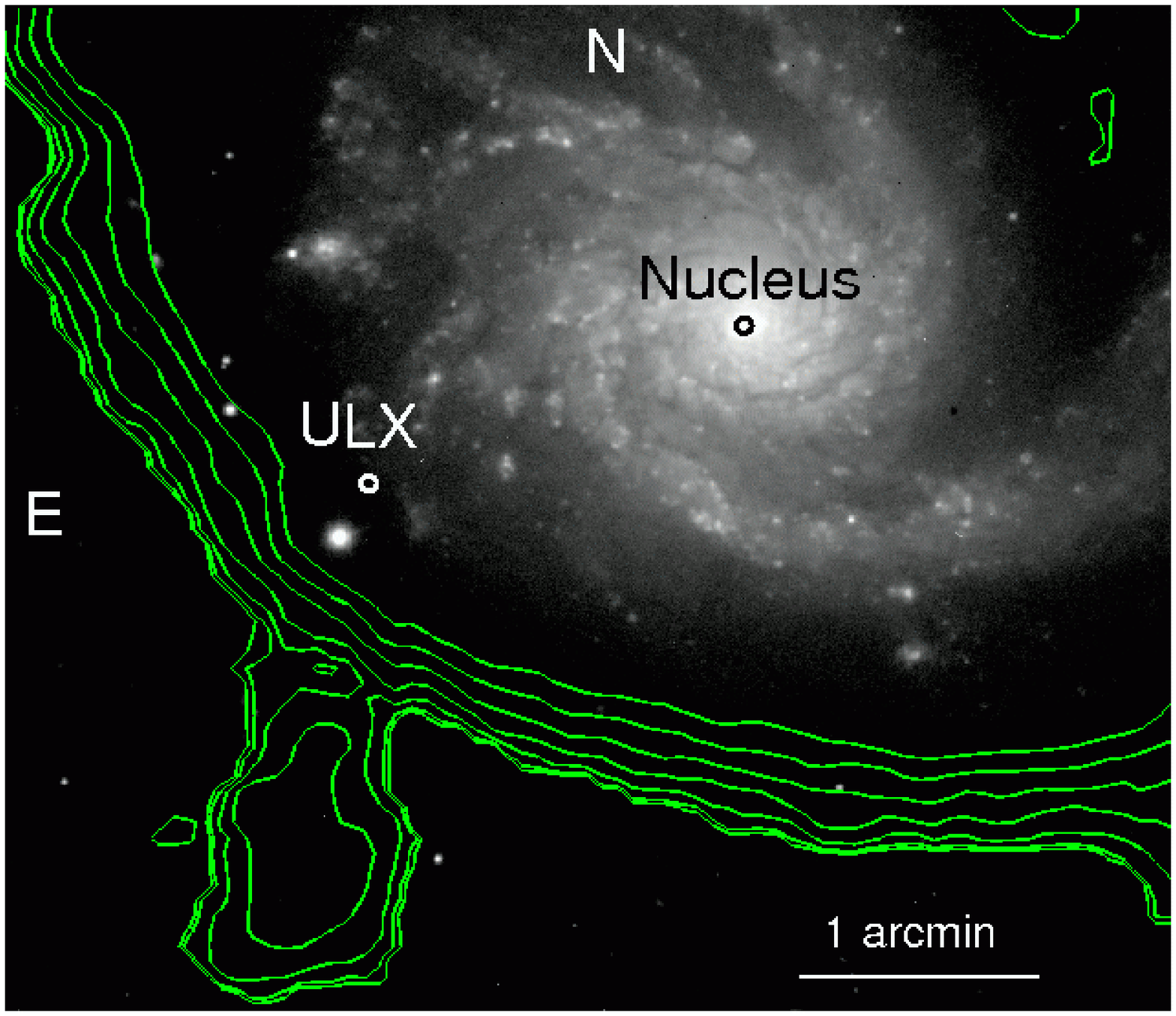, width=8.5cm, angle=0}
\caption{Top panel: H\,I 21-cm radio flux, 
represented in greyscale over the galactic disk, 
and as contours for the outer (less dense) regions. 
Contour levels are on a square-root scale from 
0.02 to 0.3 Jy km s$^{-1}$ beam$^{-1}$.
Bottom panel: the same H\,I radio flux contours 
overplotted over a {\it VLT}/FORS1 $R$-band image.}
\end{figure}

\begin{figure}
\epsfig{figure=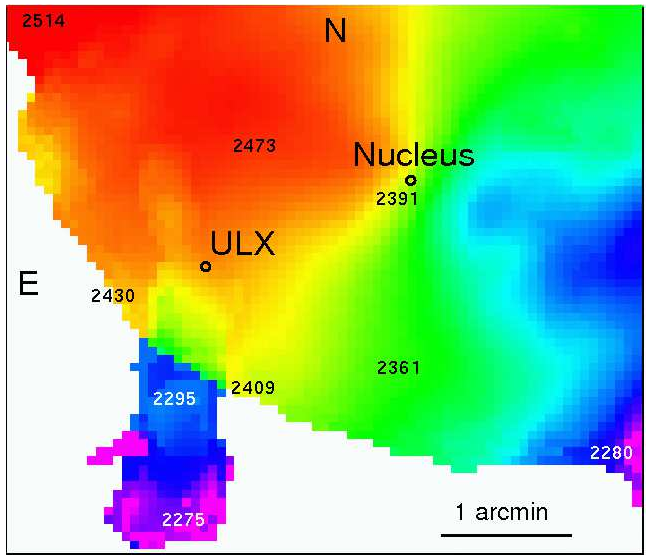, width=8.5cm, angle=0}
\caption{Velocity map of the H\,I 21-cm radio 
emission. Characteristic radial velocities (in km s$^{-1}$) are also 
marked on the plot, at various places on the disk and the cloud.
The gas cloud is consistent with tidal debris infalling onto 
the galactic disk from behind (with respect 
to our viewpoint), at a projected relative speed $> 100$ km s$^{-1}$; 
it appears as though it is impacting or merging with 
the disk at or near the ULX position.}
\end{figure}

\section{Discussion}

\subsection{Hard state of ULXs}

With an X-ray luminosity $\approx 2 \times 10^{40}$ 
erg s$^{-1}$, the ULX in M\,99 is among the brightest 
of this mysterious class of sources: the cut-off 
in the luminosity function is at $\approx 3 \times 10^{40}$ 
erg s$^{-1}$ (Gilfanov, Grimm \& Sunyaev 2004; Swartz et al.~2004).
If the emission is isotropic and Eddington-limited, 
the mass of the accreting BH is $\ga 100 M_{\odot}$.
Apart from its extreme brightness, there is a notable,  
unexplained feature associated with this source 
and other bright ULXs: the X-ray spectrum is well fitted 
by a simple power-law 
with a rather hard photon index ($\Gamma = 1.7 \pm 0.1$).
In stellar-mass BH binaries, such spectra are characteristic 
of the low/hard state (e.g., McClintock \& Remillard 2006), 
with typical X-ray luminosities 
in the standard {\it XMM-Newton} 
or {\it Chandra} bands less than a few per cent 
of the Eddington luminosity. At higher mass accretion 
rates and luminosities, 
stellar-mass BH binaries are dominated by either 
an accretion disk component (thermal, or high/soft state) or 
by a steep power-law component ($\Gamma \ga 2.5$). 
If ULXs follow the same spectral-state classification 
as stellar-mass systems, the inferred spectral slope and luminosity 
of the source in M\,99 would suggest an accreting BH 
with a mass $\sim 10^4 M_\odot$. This would require so-far 
untested formation processes, such as primordial 
remnants or nuclei of accreted dwarf galaxies 
that have recently captured a donor star.
Alternatively, ULXs may simply have a different spectral 
behaviour; for example, they might be found in 
a steady high/hard state which is not known 
in Galactic X-ray binaries. For other examples 
of bright ULXs with relatively hard power-law-like 
spectra, see Winter et al.~(2005), and Stobbart, 
Roberts \& Wilms~(2006); see also the discussion 
in Gon\c{c}alves \& Soria (2006) 
and references therein.

The underlying problem is how to explain the lack 
of evidence or the relative weakness of the accretion 
disk component in such bright sources. If the disk 
is faint because it is truncated (as in the standard low/hard state), 
we expect a low radiative efficiency, as well as a dimensionless 
accretion rate much below Eddington; therefore, 
a very high accretion rate (in physical units) 
and an even higher BH mass are needed 
to explain the observed luminosity. If the disk 
extends to the innermost stable circular orbit, 
the efficiency may be higher (standard efficiency $\sim 0.1$);
however, most of the detected photons must have been 
reprocessed in a comptonizing medium, 
because we do not see the thermal disk spectrum directly. 
This requires that most of the gravitational energy 
is released in or transferred to the upscattering 
medium, for example via magnetic coupling (Kuncic \& Bicknell 2004).
Finally, especially for a pure power-law source, 
we cannot rule out that the X-ray emission comes 
from a relativistic jet pointing towards 
our line of sight (microblazar) (K\"{o}rding, Falcke \& 
Markoff 2002). However, physical and statistical 
arguments make the strong-beaming scenario rather 
implausible as a general explanation for the whole 
population of ULXs (Davis \& Mushotzky 2004). 
Direct evidence against beamed emission is provided 
in some cases by the energy requirements of 
a photoionized nebula surrounding a ULX 
(Pakull \& Mirioni 2003). Relativistic beaming has also 
been ruled out, in one case, by the presence of quasi-periodic 
oscillations (M\,82 X-1: Strohmayer \& 
Mushotzky 2003), and in another case by the radio/X-ray flux ratio 
(Holmberg IX X-1: Miller, Fabian \& Miller 2004).

\subsection{ULXs and galaxy collisions}

Bright ULXs are preferentially associated 
with tidally interacting or collisional systems, 
or actively star-forming galaxies, or both 
(e.g., Swartz et al.~2004 for a population study).
For example, many bright ULXs have been found 
in the Antennae (Zezas \& Fabbiano 2002), 
the Cartwheel (Gao et al.~2003),
the Mice (Read 2003), NGC\,7714/15 (Smith, Struck 
\& Nowak 2005), NGC\,4485/90 (Roberts et al.~2002) 
the M\,81/M\,82 group (in M\,82, 
Holmberg II and Holmberg IX).

The simplest explanation is that galaxy collisions
trigger or enhance star formation, which, in turn, 
increases the birth rate of high-mass X-ray binaries 
(hence, the normalization of the point-source 
X-ray luminosity function). This also increases 
the probability of forming very luminous sources, 
near the upper cut-off. For example, it explains 
the large number of X-ray sources with 
luminosities $\ga 10^{39}$ erg s$^{-1}$ 
in the Antennae (Zezas \& Fabbiano 2002).

While this is probably a correct argument, 
it cannot be the whole story, especially for the small 
sample of ULXs with luminosities $\ga 10^{40}$ erg s$^{-1}$.
Some of them (e.g., Holmberg II X-1: Dewangan et al.~2004; 
Holmberg IX X-1: Miller et al.~2004) 
are in dwarf galaxies with relatively small star-formation rates. 
Others, such as the ULX in M\,99 or the brightest ULXs in NGC\,7714 
(Soria \& Motch 2004) and NGC\,4559 (Soria et al.~2005), 
are in strongly star-forming galaxies but 
far away from the main star-forming regions.
There are no ULXs in the inner disk of M\,99 
despite an SFR $\sim 10 M_{\odot}$ yr$^{-1}$;
instead, the bright ULX object of our study is at the outer edge 
of the stellar disk, where the star formation rate 
is and probably has always been orders of magnitude lower.
And yet, the ULX does seem to be associated 
with a collisional event between the disk and 
an infalling gas cloud (Section 3). How can we make sense 
of this apparently contrasting evidence?
Here, we try to speculate one possible scenario for ULX 
formation, consistent with the sketchy observational 
evidence available so far. The following statements 
are conjectures to be tested with further 
observational and theoretical modelling. 

\begin{itemize}
\item[{\it i)}] Most ULXs are not strongly beamed. They are powered 
by BHs with masses $\sim 20$--$200 M_{\odot}$; the upper limit 
can easily be reduced to $\la 100 M_{\odot}$ if mild 
anisotropy or mild super-Eddington emission are allowed.
The donor is likely to be a Roche-lobe-filling OB donor star, 
which can supply the required amount of accreting gas 
over its nuclear timescale (Rappaport, Podsiadlowski \& Pfahl 2005; 
Copperwheat et al.~2006).
If so, ULXs represent the upper end of high-mass X-ray binaries, 
consistent with their preferential location in young 
stellar environments; the accreting BHs are a factor 
of a few more massive than Galactic systems, 
and require stellar progenitors with masses 
$\sim 50$--$400 M_{\odot}$. 

\item[{\it ii)}] A starburst or high SFR may favour but 
is not a sufficient 
condition for the formation of bright ULXs; other conditions 
may be required. For example, low metal abundance 
is essential to allow the evolution of a very massive star 
into a massive BH (Pakull \& Mirioni 2003; Heger et al.~2003). 
A high SFR is not a necessary condition, 
either, as proved by the occasional presence of ULXs in 
gas-rich but low-SFR environments. Super star-clusters are 
also not a necessary condition for ULX formation.

\item[{\it iii)}]Galaxy-galaxy or cloud-galaxy collisions may induce 
the formation of bright ULXs, as well as trigger intense star 
formation, as separate, parallel consequences. This could explain 
why the two phenomena (ULXs and starbursts) are often, 
but not always, associated with each other. For example, 
local collisional events in NGC\,4559 (Soria et al.~2005) 
and M\,99 may be responsible for their respective ULXs 
in metal-poor regions at the very edge of their stellar disks.

\end{itemize}

The key issue we need to explain is why a collisional event 
would {\it directly} (i.e., not simply through 
enhanced star formation) favour the formation 
of a very massive star, which may then become a ULX 
progenitor if other conditions (e.g., low metal abundance) 
are also satisfied.
A possible explanation is suggested by studies 
of the star-formation process in the Galactic 
protocluster NGC\,2264-C, in the Cone Nebula 
(Peretto, Andr\'{e} \& Belloche 2006). In that case, 
a molecular clump with a mass $\approx 1700 M_{\odot}$ 
is undergoing a large-scale collapse onto its central region, 
on a dynamical timescale (essentially in free fall). This results 
in the formation of a few massive protostars, one of which 
has already reached a mass $\approx 40 M_{\odot}$ but 
is still accreting at a rate $\sim 10^{-3} M_{\odot}$ yr$^{-1}$ 
(Peretto et al.~2006).
It is plausible that two or three protostars would accrete 
a few $100 M_{\odot}$ of gas and coalesce even before 
ending their protostellar phase ($\approx 3 \times 10^5$ yr).
Based on this scenario, we speculate that the formation 
of a star with an initial mass of a few $100 M_{\odot}$ 
may occur in a protocluster such as NGC\,2264-C, 
via massive global gas infall and mergers already 
during the protostellar phase (Soria 2006); this is 
perhaps a more common process than the runaway merger 
of O stars in the core of super star-clusters 
(Portegies Zwart \& McMillan 2002; 
Freitag, G\"{u}rkan \& Rasio 2006). 
%For example, this may have been the origin of the Pistol star 
%($M \approx 200 M_{\odot}$) **REF**

It can be shown that an external shock or pressure wave  
can trigger the global, dynamical collapse of a 
molecular clump such as NGC\,2264-C; instead, an isolated clump 
in hydrostatic equilibrium would have fragmented 
into much smaller Jeans-mass protostars (for recent 
reviews, see for example Struck 2005, Elmegreen 2002, 2004).
%astro-ph/0405552
The critical momentum required to trigger 
the collapse of a molecular clump 
is of order of the clump mass times its sound speed; 
moreover, the accretion rate 
onto the protostellar cores is proportional 
to the momentum imparted to the collapsing 
molecular clump (Motoyama \& Yoshida 2003).
Supernova shocks are a possible trigger of dynamical 
collapses in nearby molecular clumps; a fast, 
colliding gas cloud may carry even more energy 
and momentum than a supernova.
Thus, we speculate that collisional events 
(such as the cloud-disk collision at the outer edge
of M\,99, or the possible dwarf galaxy-disk collision 
in NGC\,4559: Soria et al.~2005) may lead to the formation 
of extremely massive protostars, via dynamical collapse 
and protostellar mergers on a timescale $\sim 10^5$ yr.

The second step of the process to form a ULX would 
then be to ensure that the massive stellar progenitor 
retains most of its gas until 
core collapse. We speculate that this is where 
low metal abundance comes into play, because 
it strongly reduces the mass loss rate through 
line-driven stellar winds (e.g., Eldridge \& Vink 2006, 
and references therein) and therefore leaves behind 
a bigger BH. This is why 
the (metal-rich) Pistol star in the Quintuplet Cluster 
(initial mass $\approx 200$--$250 M_{\odot}$: Figer et al.~1998), 
the Wolf-Rayet stars in the Arches cluster, 
or the massive protostars in the NGC\,2264-C protocluster
will never form massive BHs, and perhaps why there are 
no ULXs in that part of our Galaxy.
Conversely, we expect metal abundance to be much lower 
at the outer edge of the disk in M\,99, perhaps 
made even lower by the infalling gas clouds 
(see, e.g., Chen, Hou \& Wang 2003, and Andrievsky et al.~2003, 
for recent studies of metal abundances as a function 
of galactocentric distance in the Milky Way; and MacArthur 
et al.~2004 for similar trends in other disk galaxies).

In summary, our suggested scenario differs from 
models based on the runaway coalescence of O stars 
in super star-clusters, because it does not 
require such massive systems. We speculate 
that the formation of a massive stellar progenitor 
may occur in smaller ($\sim 10^3$--$10^4 M_{\odot}$) 
protoclusters, during their embedded phase, through 
infall and merger processes similar to what is currently 
observed in NGC\,2264-C. Protoclusters of that size 
are massive enough to produce O stars, but would 
by no means be classified as super clusters; in fact, 
they may not even be massive enough to survive 
their embedded phase as bound systems (Kroupa \& Boily 2002).

Our scenario also differs from ULX formation models 
based on Population-III stars, because it only requires 
low but not primordial 
abundances, and therefore it can work at any redshift. 
At zero metallicity, stars with initial masses 
$\approx 140$--$300 M_{\odot}$ 
may be disrupted by pair-instability supernovae that 
leave no remnants (Heger et al.~2003). Instead, 
at $Z \sim 0.1 Z_{\odot}$, stars in that mass range 
may become the progenitors of accreting BHs in ULXs, 
via direct collapse, precisely 
in the mass range required by the X-ray observations.
Further discussion of each of those speculations, 
and detailed comparisons with the observations, is beyond 
the scope of this paper.

\section{Conclusions}

We studied the X-ray properties of M\,99, the most massive 
spiral galaxy in the Virgo Cluster. As expected 
from its high SFR ($\approx 10 M_{\odot}$ yr$^{-1}$), 
the X-ray emission is dominated by a soft thermal-plasma 
component. The total unabsorbed luminosity (not including 
a bright ULX) inside the $D_{25}$ ellipse  
is $\approx (1.2 \pm 0.2) \times 10^{41}$ erg s$^{-1}$ 
in the $0.3$--$12$ keV band; about $15\%$ of this is due 
to resolved or unresolved discrete sources. 
The emission appears almost uniformly diffused 
across the inner disk (at $\la 5$ kpc from the nucleus), 
and there is no starburst core. The temperature 
of the hot gas, $kT \la 0.30$ keV, is also rather low, 
typical of disk emission rather 
than of a starburst environment. 

We briefly discussed the morphology of the nucleus 
in the optical and UV bands, showing the presence 
of a (redder) massive nuclear star cluster 
and a few smaller, much bluer clusters of young stars 
around it. The spatial resolution 
of {\it XMM-Newton} does not reveal whether there are 
faint point-like X-ray sources in the nuclear region, 
and if so, whether they are located in the old 
nuclear cluster or associated to the young stars 
around it.

The main goal of this paper was a study of the properties and origin 
of a bright ULX at the outer edge of the stellar disk.
Its unabsorbed luminosity of $\approx 2 \times 10^{40}$ erg 
s$^{-1}$ in the $0.3$--$12$ keV band, together with 
a pure power-law spectrum with photon index $\Gamma \approx 1.7$, 
are difficult to interpret in the framework 
of classical spectral states for stellar-mass BHs.
Such high/hard states are not uncommon in ULXs 
but are not generally seen in Galactic BHs.
It suggests that we are not seeing the accretion disk 
directly, and that most of the gravitational power 
is efficiently transferred to a comptonizing region.

An intriguing new discovery is that there is a massive gas cloud 
(H\,{\footnotesize I} mass $\sim 10^7 M_{\odot}$) 
seen in projection very close to the ULX position. From 
its radial velocity, we suggested that this cloud 
is falling onto the galactic disk (from behind) and 
perhaps impacting it near the ULX location.
The cloud occupies $\approx 3\%$ of the projected area 
of the stellar disk; therefore, we cannot entirely 
rule out the possibility that it is a chance association. 
However, we explored the speculative idea that there is a direct 
connection between collisional events and ULX formation, 
and not simply an indirect effect due to an enhancement 
in star formation. We suggested a possible scenario 
of ULX formation that would be consistent with this 
interpretation. We argued that cloud collisions 
may trigger a large-scale dynamical collapse of molecular clumps, 
rather than Jeans-mass fragmentation; in turn, 
this may lead to the formation of very massive stellar 
progenitors in the cluster core. If the metal abundance 
is low ($Z \sim 0.1 Z_{\odot}$,  not primordial 
abundances), we speculate that 
the star  might retain enough 
of its gas to directly-collapse into a BH massive enough 
(up to $\sim 100 M_{\odot}$) to explain the luminosity 
of the brightest ULXs.

This scenario has to be tested with further individual 
and population studies of ULXs. At the individual level, 
the suggestion that ULXs have masses up to an order of magnitude 
higher than Galactic BHs (but not more) needs to be tested 
with more advanced X-ray spectral modelling, 
and a better understanding of the thermal disk 
and comptonized emission components in such systems.
At the population level, we need a more systematic 
study of the spatial association between ULXs, low-metallicity 
environments and collisional events. Further work 
on this issue is currenly under way (D. Swartz et al., 
in preparation).

\section*{Acknowledgments}
We thank Manfred Pakull for discussions on the nature of ULXs 
and particularly his intuition about the role of metal abundance. 
We thank Mark Cropper for discussions on the effect  
of cloud and satellite collisions. We are also grateful  
to Alex Filippenko, Anabela Gon\c{c}alves, Zdenka Kuncic, 
Dave Pooley and Kinwah Wu. RS acknowledges support 
from an OIF Marie Curie Fellowship. DSW is grateful 
for financial support under Chandra grant G05-6092A 
to D. Pooley and A. V. Filippenko.
%, as well as financial support 
%from the University of Sydney during his visit there.

\end{document}